# Broadband tunable microwave photonic radar for simultaneous detection of human respiration, heartbeat, and speech with deep learning-based speech recognition

**Lei Gao[†], Dingding Liang[†], Jiawei Gao[†], Chulun Lin, Zhiqiang Huang, Taixia Shi, and Yang Chen[*]**

East China Normal University, School of Communication and Electronic Engineering, Shanghai Key Laboratory of Multidimensional Information Processing, Shanghai, China, 200241

† These authors contributed equally to this paper

**Abstract**: Multimodal vital sign monitoring and speech detection hold significant importance in medical health, public safety, and other fields. This study proposes a broadband tunable microwave photonic radar system that can simultaneously monitor respiration, heartbeat, and speech. The system works by generating broadband radar signals to detect subtle skin displacements caused by these physiological activities. It then utilizes phase variations in radar echo signals to extract and reconstruct the corresponding physiological signals. In order to enhance the processing capability for speech signals, a convolutional neural network with a dual-channel feature fusion model is incorporated, enabling high-precision speech recognition. In addition, the system's frequency-tunable characteristic allows it to flexibly switch frequency bands to adapt to different working environments, greatly improving its practicality and environmental adaptability. In concept-verification experiments, speech signals were reconstructed and recognized in the Ku, K, and Ka bands, achieving recognition accuracies of 97.20%, 98.07%, and 97.43%, respectively. The system's capability to detect multimodal vital signs was also thoroughly validated using a respiratory and heartbeat simulator. During a 20-second monitoring period, while accurately reconstructing speech, the maximum average error counts for respiratory and heartbeat monitoring were 0.39 and 0.87, respectively, proving its reliability and effectiveness in multimodal vital sign monitoring.

**\*Seventh author, E-mail: ychen@ce.ecnu.edu.cn

## 1 Introduction

With the rapid development of artificial intelligence and Internet of Things (IoT) technology, accurate detection of human voice signals and vital signs such as respiration and heartbeat has become a crucial demand in many application scenarios[1-3]. For example, as shown in Fig. 1(a), some patients with neurological disorders, such as Parkinson's disease and amyotrophic lateral sclerosis, have limited mobility and lose their voice as the disease progresses[4,5]. Therefore, monitoring the vital signs and voice signals of patients with neurological disorders enables the timely detection of subtle disease progression. Such monitoring facilitates prompt treatment regimen adjustments, slowing down the extent of disease progression and preventing complications. Figure 1(b) shows that monitoring the vital signs and speech signals of trapped individuals is crucial for accurate damage classification and timely rescue at disaster sites such as earthquakes[6]. In the war, the detection of vital signs is crucial for battlefield medical decisions for friend forces[7]. However, from an adversarial perspective, real-time detection of the heartbeat, respiration, and speech signals of enemy personnel in the surrounding environment can offer more comprehensive intelligence on the adversaries, which is beneficial for our tactical planning, as depicted in Fig. 1(c). To accurately monitor these vital signs, many feasible solutions have emerged, which can be divided into contact and contactless detection technologies[8].

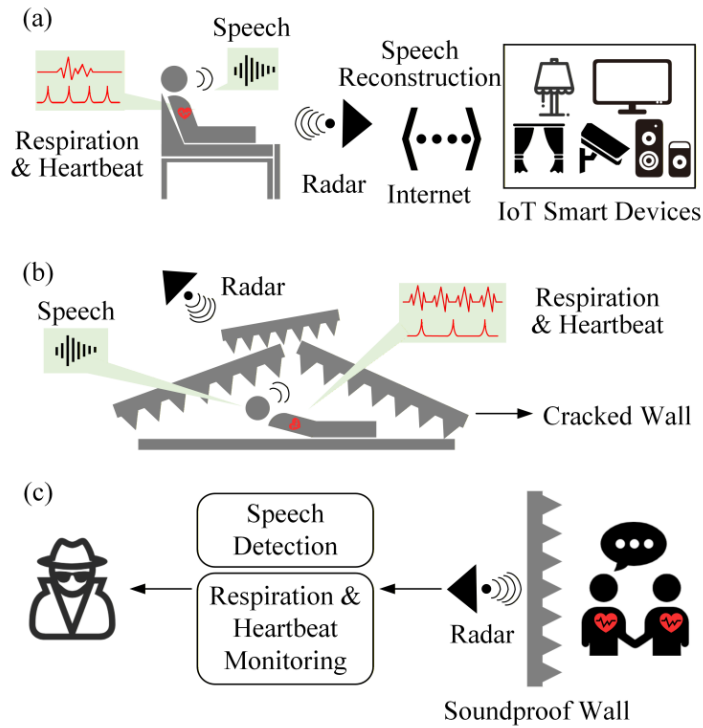

**Fig. 1** Applications of the vital sign monitoring and speech detection in different scenarios. (a) Medical monitoring for patients in clinical environments. (b) Rescue operations at large-scale disaster sites. (c) Eavesdropping and detection in modern warfare.

In the medical field, traditional monitoring techniques have relied on contact-based methods, such as photoplethysmography[9], pulse oximetry[10], electrocardiography[11], and skin-attachable acoustic sensors[12]. However, in various situations, such as in the case of infected patients or of patients suffering from mental illness or affected by severe burns or injuries, the use of contact sensors is not possible or recommended[13], because it may cause not only discomfort and irritation on the skin but also transmission of contagious diseases[14]. In such cases, the use of contactless monitoring devices, such as radar systems, can help healthcare professionals by providing critical information about the patient's state[15].

Radar technology, as a contactless detection method, has shown great potential in the fields of vital sign monitoring and speech perception. In terms of vital signs monitoring, the idea of remotely monitoring physiologic functions in humans via radars started as early as the 1970s[16,17]. However, the development was limited due to the cumbersome and expensive apparatus used in those years. Subsequently, advancements in technology allowed the development of cheap and compact radar systems mainly based on continuous wave (CW)[18], ultrawideband (UWB)[19], or frequency modulated continuous wave (FMCW)[20] techniques. Furthermore, modern radar technology has not only enabled contactless monitoring of vital signs but also been extended to the detection of life activity states, such as posture[21] and sleep status[22]. As a contactless monitoring method, radar can perform long-distance, interference-free monitoring of vital signs, which provides patients with a certain degree of freedom and reduces the risk of infectious disease transmission, offering a safer and more comfortable monitoring experience[23]. In the field of sound perception, the earliest application involved using radar to measure the movement of the tracheal posterior wall during speech, thereby achieving speech perception[24]. This demonstrated the feasibility of radar in contactless speech detection. Subsequently, researchers have explored the use of millimeter-wave radar for contactless speech detection, which demonstrated that the millimeter-wave radar can capture high-quality speech signals and enable accurate speech recognition[25,26]. However, when noise and other interferences occur in the radar operating frequency band, traditional radar-based methods for vital signs monitoring and speech perception struggle to perform adaptive detection across multiple frequency bands and are susceptible to electromagnetic interference. Therefore, there is an urgent need for new methods that can break through these limitations.

Microwave photonic radar, which possesses unique advantages in addressing the aforementioned issues, has received widespread attention and research. In the transmitter of the microwave photonic radar, high-frequency and broadband radar signals can be generated through photonic frequency multiplication[27], optical injection[28], acousto-optic frequency shift loops[29], etc. In the receiver, photonic de-chirping is employed to convert high-frequency broadband signals into low-frequency narrowband signals[30]. Currently, research on microwave photonic radar primarily focuses on high-resolution target range measurement and imaging, with relatively fewer studies on vital signs monitoring. Recently, a research team has demonstrated a contactless vital signs monitoring system based on microwave photonic radar[31]. By using a frequency shift loop, they generated a stepped-frequency radar signal with an instantaneous bandwidth of 10 GHz and extracted the respiration of a breathing simulator and a toad using microwave photonic signal processing. The monitored respiratory rates of the two breathing simulators were 12 rpm and 16.5 rpm, respectively. Both radar and camera were used to monitor the respiration of the toad, with a cross-correlation coefficient of 0.746 between the two monitoring methods. We have also previously realized contact and contactless monitoring of human respiration and heartbeat[32]. By analyzing the output optical intensity of a fiber Bragg grating (FBG) and the phase of the radar de-chirped signal, simultaneous contact and contactless monitoring of respiration and heartbeat have been achieved. Vital signs of three individuals were simultaneously obtained via contact and contactless methods, with the maximum measurement errors for respiratory rate and heartbeat rate being 1.6 rpm and 2.3 bpm, respectively. When the vital signs monitoring time is reduced to 5 seconds, there is no significant change in the measurement errors of respiratory rate and heartbeat rate. In the field of speech perception, to the best of our knowledge, there have been no reported studies or applications of microwave photonic radar.

In this work, we propose and experimentally demonstrate a broadband tunable microwave photonic radar system capable of simultaneously monitoring speech, respiration, and heartbeat across the Ku, K, and Ka bands. For speech detection, by employing a dual-channel feature fusion convolutional neural network (CNN) model incorporating both spectrogram and Mel-spectrogram for training, the system achieves speech recognition accuracies of 97.20% in Ku band, 98.07% in K band, and 97.43% in Ka band. For respiratory and heartbeat monitoring, the maximum average absolute error counts for respiration and heartbeat over a 20-second monitoring time are 0.36 and 0.87. The proposed system offers a promising solution for a variety of scenarios, including monitoring the physiological activities of patients during medical treatment, locating and rescuing trapped individuals in disaster situations, and gathering vital signs information from adversaries on the battlefield to inform tactical planning. In addition, owing to the utilization of microwave photonic technology, this system holds the potential for adaptively selecting different frequency bands. This enables it to better adapt to various application scenarios, facilitating more effective speech signal reconstruction as well as respiratory and heartbeat monitoring.

## 2    Principle

*2.1 Feasibility Study*

The central mechanism of human vocalization is largely based on respiratory airflow-driven vocal fold vibration, which generates specific acoustic waveforms by resonating with the vocal tract based on laryngeal movements. This vibration triggers a micrometer to millimeter scale periodic displacement of the laryngeal skin, which corresponds strictly to acoustic parameters such as speech frequency and intensity. As shown in Fig. 2, a millimeter-wave radar detects minute skin displacements by transmitting high-frequency electromagnetic waves, receiving the echo signal reflected from the laryngeal skin, and then demodulating the phase change of the signal. When the laryngeal skin is displaced, the phase change of the echo signal is proportional to the vocal fold vibration. Hence, by detecting minute vibrations of the vocal fold, the speech signal can be reconstructed. Similarly, the millimeter-wave radar can detect minute vibrations associated with respiration and heartbeat, thereby enabling the monitoring of these vital signs. Therefore, the technical feasibility of this study is based on the synergy between the high sensitivity of millimeter waves for detecting minute displacements and the biomechanics of vocal fold vibration in the human vocalization, chest fluctuation in respiration and heartbeat.

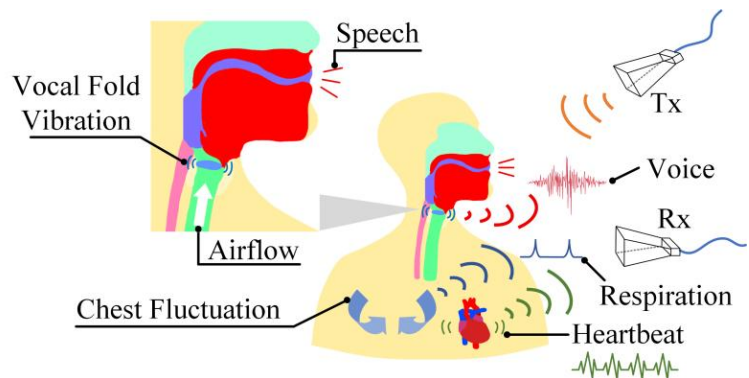

**Fig. 2** Schematic diagram of feasibility study.

## 2.2 Principle

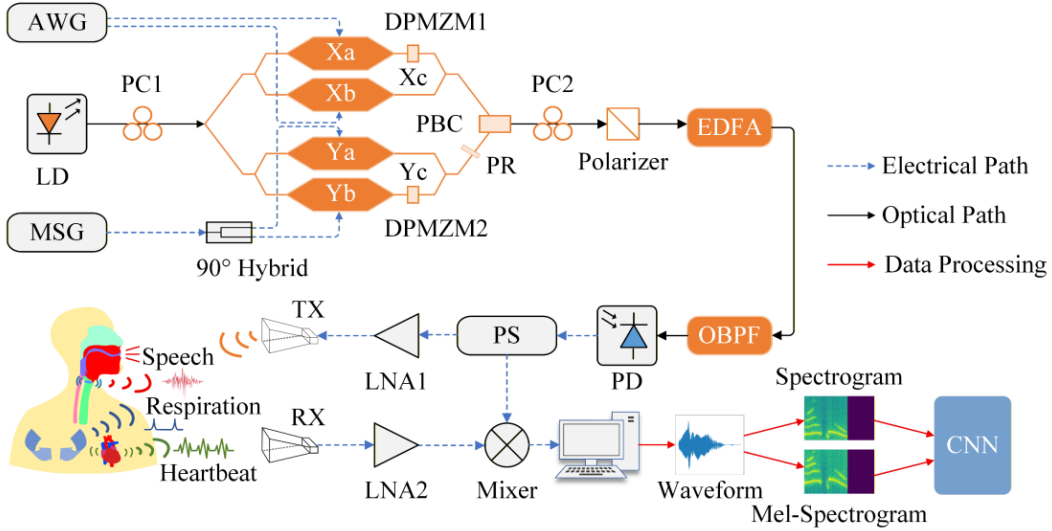

**Fig. 3** Schematic diagram of the microwave photonic radar system for simultaneous detection of human respiration, heartbeat, and speech. LD, laser diode; PC, polarization controller; DPMZM, dual-parallel Mach–Zehnder modulator; PR, polarization rotator; PBC, polarization beam combiner; EDFA, erbium-doped fiber amplifier; AWG, arbitrary waveform generator; MSG, microwave signal generator; LNA, low-noise amplifier; OBPF, optical band-pass filter; PD, photodetector; PS, power splitter; CNN, convolutional neural network; Tx, transmitting antenna; Rx, receiving antenna.

The schematic diagram of the proposed microwave photonic radar system for monitoring human vital signs and detecting human speech signal is shown in Fig. 3. A CW light wave generated by a laser diode (LD) is sent to a dual-polarization dual-parallel Mach−Zehnder modulator (Dpol-DPMZM) via a polarization controller (PC1). The CW light wave as the optical carrier can be expressed as

$$E_c(t) = E_c \exp(j2\pi f_c t), \qquad (1)$$

where $E_c$ and $f_c$ are the amplitude and frequency of the optical carrier, respectively. The Dpol-DPMZM consists of two dual-parallel MZMs (DPMZMs). DPMZM1 is driven by two phase-orthogonal intermediate frequency (IF) linearly frequency-modulated (LFM) signals generated from an arbitrary waveform generator (AWG). One of the LFM signals can be expressed as

$$V_{IF}(t) = V_{IF} \cos\left[2\pi\left(f_0 t + \frac{1}{2}kt^2\right)\right] \quad t \in [0,T], \tag{2}$$

where $V_{IF}$, $f_0$, and $k$ denote the amplitude, initial frequency, and chirp rate of the LFM signal. The instantaneous frequency of the LFM signal can be described as: $f_{IF}=f_0+kt$. The two sub-MZMs (Xa and Xb) of DPMZM1 are biased at the minimum transmission point (MITP), and the main MZM (Xc) is biased at the quadrature transmission point (QTP). Under the small-signal modulation condition, DPMZM1 outputs a carrier-suppressed single-sideband (CS-SSB) signal corresponding to this LFM signal. Specifically, it generates the −1st-order modulation optical sideband with a center frequency of $f_c-f_{IF}$. The optical signal output from DPMZM1 can be written as

$$E_{DPMZM1}(t) \propto E_1 \exp\left[j2\pi(f_c - f_0)t - j\pi kt^2\right]. \tag{3}$$

The two RF ports of DPMZM2 are connected to a microwave signal generator (MSG) via a 90° electrical hybrid coupler. A single-tone microwave signal centered at $f_s$ generated from the MSG is applied to the hybrid coupler and then sent to DPMZM2. The operating states of the two sub-MZMs (Ya and Yb) and the main MZM (Yc) of DPMZM2 are the same as those of DPMZM1. Under these circumstances, DPMZM2 is also operated as a CS-SSB modulator that outputs an optical signal with +1st-order modulation optical sideband corresponding to this single-tone signal. The optical signal output from DPMZM2 can be written as

$$E_{DPMZM2}(t) \propto E_2 \exp\left[j2\pi(f_c + f_s)t\right], \tag{4}$$

where $E_2$ is the amplitude of the optical signal. Thanks to a 90° polarization rotator on the branch where DPMZM2 is located, the Dpol-DPMZM is capable of generating two CS-SSB optical signals with orthogonal polarization states. Then, these two optical signals are combined through the built-in polarization beam combiner (PBC) in the Dpol-DPMZM. Following the Dpol-DPMZM, the output optical signal is fed into a polarizer via PC2. By adjusting PC2, one of the orthogonal optical polarization states output from the Dpol-DPMZM is oriented at a 45° angle relative to the principal axis of the polarizer, thereby enabling the orthogonally polarized optical signals to be projected onto this axis. After the polarizer, an erbium-doped fiber amplifier (EDFA) is utilized to boost the optical signal power. Then an optical band-pass filter (OBPF) is used to eliminate the out-of-band amplifier spontaneous emission noise introduced by the EDFA. After the OBPF, the optical signal is sent to a photodetector (PD) for optical-to-electrical conversion. The electrical LFM signal generated from the PD is separated into two paths by a power splitter (PS): one output of the power splitter is connected to a transmitting antenna via a low-noise amplifier (LNA1); the other output of the power splitter is sent to a mixer as a local oscillator (LO) signal. The transmitted signal $i_{\text{Trans}}$ can be written as

$$i_{\text{Trans}}(t) \propto E_{\text{T}} \cos\left[2\pi\left(f_s t + f_0 t + \frac{1}{2}kt^2\right)\right], \tag{5}$$

where $E_{\text{T}}$ is the amplitude of the transmitted signal. Notably, the frequency band of the transmitted signal can be tuned over a wide range by adjusting the frequency of the single-tone signal. After power amplification, the transmitted signal is sent to free space for human speech and vital sign monitoring through the transmitting antenna.

At the receiving end, the radar echoes reflected from the human body are received by a receiving antenna with a delay time $\Delta\tau$, which can be written as:

$$i_{\text{Echo}}(t) \propto E_R \cos\left\{2\pi\left[f_s(t-\Delta\tau) + f_0(t-\Delta\tau) + \frac{1}{2}k(t-\Delta\tau)^2\right]\right\}, \tag{6}$$

where $E_R$ is the amplitude of the echo signal. The echo signal is amplified by another LNA (LNA2) and mixed with the LO signal in the mixer. The de-chirped signal from the mixer is sampled to extract the human speech and vital sign information. The de-chirped signal can be expressed as

$$i_{\text{de-chirp}}(t) \propto \cos\left[2\pi\left(f_0\Delta\tau + f_s\Delta\tau + kt\Delta\tau - \frac{1}{2}k\Delta\tau^2\right)\right], \tag{7}$$

Since $k\Delta\tau^2/2$ is far less than $(f_0+f_s)\Delta\tau$, $k\Delta\tau^2/2$ in Eq. (7) can be ignored. In the case of contactless monitoring, the frequency and phase of the radar de-chirp signal can be expressed as:

$$f_{\text{de-chirp}} = k\Delta\tau, \tag{8}$$

$$\varphi_{\text{de-chirp}}(t) = 2\pi(f_0+f_s)\Delta\tau = \frac{4\pi[R_0+x(t)]}{\lambda}, \tag{9}$$

where $\Delta\tau = 2[R_0+x(t)]/c$, $\lambda = c/(f_0+f_s)$. $R_0$ is the range between the radar antenna pair and the human body, $x(t)$ is the skin displacements produced by speech, respiration, and heartbeat, and $c$ is the speed of light in a vacuum.

Then, we can recover the human speech, respiration, and heartbeat information based on the phase changes of the radar de-chirped signal, with the assistance of signal processing algorithms. In addition, when there are multiple human subjects at varying distances, we can determine the distance of each individual by analyzing the distribution of frequency components along the frequency axis in the de-chirped signal. Subsequently, we can independently retrieve the phase information contributed by each individual at distinct de-chirped frequencies[31], thereby simultaneously obtaining the speech, respiration, and heartbeat information of multiple individuals.

## 2.3 Signal Processing

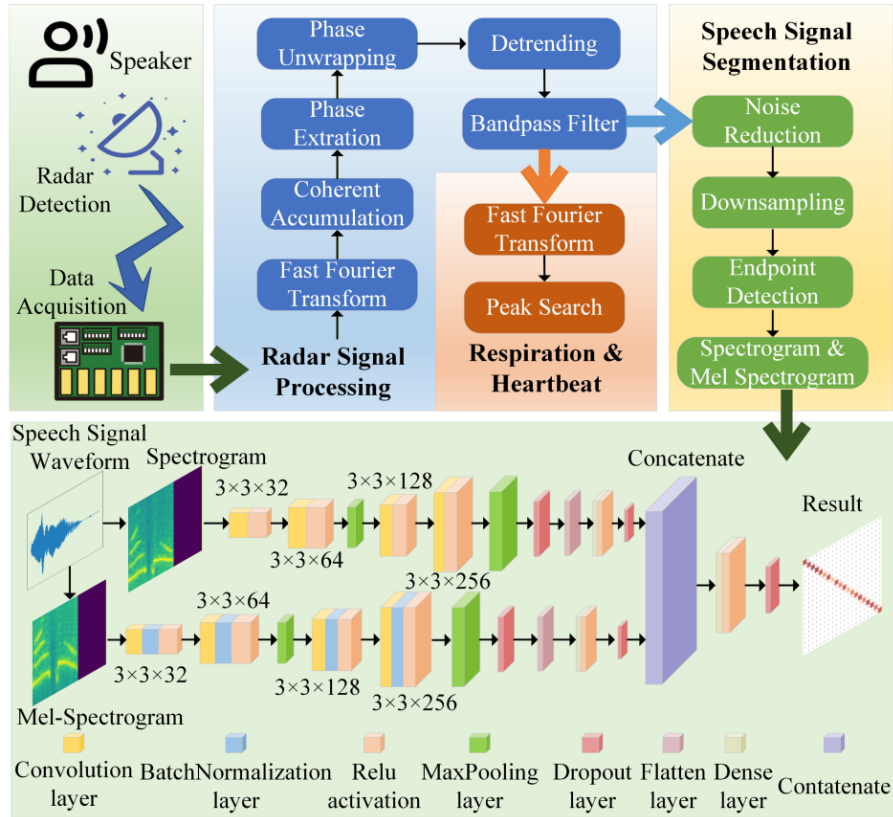

**Fig. 4** Flow chart of signal extraction and processing in contactless speech and vital sign monitoring.

The flow chart of the signal extraction and processing in contactless monitoring is shown in Fig. 4. The radar signal for monitoring speech and vital signs is received by the system and then de-chirped at the mixer. The de-chirped signal from the mixer is sampled by an oscilloscope (OSC) and converted into a digital signal, which is then transmitted to a PC for further signal processing.

In digital signal processing, the M sampling points within each signal period constitute the fast time dimension, and the echo signals of N periods are rearranged along the slow time axis to form a two-dimensional matrix. The range of the human body is obtained by performing a fast Fourier transform (FFT) of the fast time axis. When performing FFT along the fast-time axis, the extremely short period of the LFM signal allows for the omission of skin displacement caused by respiration and heartbeat. Consequently, the human body can be reasonably assumed to maintain quasi-static conditions during this transient interval. To ensure that the range bin of the human body is searched, a coherent accumulation operation is employed to improve the ranging accuracy. Considering that the skin displacements caused by speech, respiration, and heartbeat are very small, the phase detecting scheme is employed to achieve minute displacement detection with high accuracy. After the range of the human body is obtained, the phase change strictly linearly correlated with the skin displacements is extracted through phase unwrapping and detrending. Finally, three distinct digital bandpass filters are specifically designed to reconstruct speech, respiratory, and heartbeat signals, respectively, by isolating their characteristic frequency bands with tailored passband parameters. For respiratory and heartbeat monitoring, respiratory and heartbeat rates can be obtained by performing FFT and peak search on the reconstructed respiratory and heartbeat signals after filtering, respectively. For speech detection, noise reduction is performed using spectral subtraction algorithm, then the signal is resampled at 8 kHz and segmented via energy-based endpoint detection, resulting in a clean reconstructed speech signal ready for further processing.

To achieve accurate recognition of reconstructed speech signals, a dual-channel feature fusion CNN model has been further developed. This model integrates both the spectrogram and Mel-spectrogram representations of speech signals, leveraging their respective features across these two dimensions to attain improved speech recognition accuracy. To accomplish this, the reconstructed speech signals—following endpoint detection—are transformed into both a standard spectrogram and a Mel-spectrogram using the short-time Fourier transform (STFT) and a Mel filter bank, respectively. Then, these representations are fed into the two parallel branches of the dual-channel feature fusion CNN for subsequent processing. In the CNN branch for spectrogram, a network structure consisting of 2D convolution layer, rectified linear unit (ReLU) activation function, max-pooling layer, dropout layers, flatten layers, and dense layers are used for feature extraction and processing of the spectrogram of the reconstructed speech waveform. Specifically, the model utilizes four 2D convolutional layers to effectively extract discriminative features from the speech signal. After extracting features through each convolutional layer, the ReLU activation function is applied to introduce non-linearity into the resulting feature sequences, thereby improving the model's computational efficiency. Two max-pooling layers are primarily employed to perform dimensionality reduction while retaining the prominent features within local regions. Once the feature maps have gone through the second max-pooling layer for dimensionality reduction, a dropout layer is then applied. This layer randomly deactivates a certain proportion of neurons during the training process, thereby preventing the model from overfitting to the training data. Subsequently, a flatten layer is employed to reshape the multidimensional feature maps into a one-dimensional feature vector. This vector is then processed by a dense layer, followed by a ReLU activation function and another dropout layer to improve efficiency and prevent overfitting, ultimately producing preliminary classification results. Similar to the CNN branch for spectrogram, the CNN branch for Mel-spectrogram follows the same structure, but each 2D convolution layer is followed by a batch normalization layer and then a ReLU activation function. This design normalizes and adaptively adjusts the input of each layer, thereby accelerating model training, improving stability, and mitigating gradient vanishing/exploding issues. The results of the two CNN branches are finally fused in the concatenate layer and undergo anti-overfitting and regularization to generate the final speech recognition results.

*2.4 Informed Consent Statement*

In this study, where a broadband tunable microwave photonic radar was employed as a sensor to measure human respiratory, heartbeat, and voice information, a total of 10 volunteers were recruited. We obtained informed consent from each of these volunteers, specifically granting permission to record their voice information for research purposes.

It is crucial to note that during the experimental process, there was no direct exposure of the volunteers to the radar. Instead, we conducted equivalent experiments to simulate the real-world scenarios. For voice information collection, we used a loudspeaker to play the pre-recorded voice of the volunteers, and the radar was directed at the loudspeaker to simulate the situation of measuring human laryngeal vibrations. Additionally, to simulate human respiratory and heartbeat patterns, we utilized a respiratory and heartbeat simulator.

This approach ensured that the entire experimental procedure did not involve any direct irradiation of the human body by the radar. As a result, the risks associated with the study were effectively controlled. The use of these alternative methods, which bypass direct human exposure, was thoroughly reviewed and ethically approved by the Ethics Committee of East China Normal University, providing a safe and ethical framework for our research.

## 3  Experimental results and discussion

To verify the validity of the proposed microwave photonic radar system for simultaneous detection of human respiration, heartbeat, and speech, a series of experiments are carried out. Considering the substantial data requirements of deep learning algorithms and the controllability of the experiments, this study employs high-fidelity loudspeakers as the standardized sound sources to play pre-recorded human speeches. The vibrations generated by these loudspeakers are utilized to simulate the minute skin displacements caused by human speech in real-world applications. Then we systematically present the experimental results in this section from the following three aspects: 1) Broadband tunable radar signal generation; 2) Reconstruction and recognition of speech signals; and 3) Concurrent vital sign monitoring and speech detection. In addition, we finally conduct a systematic exploration and investigation into the fundamental significance and practical application value of broadband frequency tunability in microwave photonic radar for human speech reconstruction and non-contact vital sign monitoring.

*3.1 Broadband tunable radar signal generation*

In the experiment, a CW light wave emitted by the LD (ID Photonics CoBrite DX1-1-C-H01-FA), with a central wavelength of 1549.89 nm and a power of 16 dBm is injected into the Dpol-DPMZM (Fujitsu FTM7977HQA) via PC1. The AWG (Keysight M8190A) is used to generate two phase-orthogonal IF LFM signals with a central frequency of 3 GHz and a bandwidth of 2 GHz. The signal has a period of 100 μs, within which the LFM signal spans 80 μs, leaving a 20 μs interval without any signal transmission, thus yielding a duty cycle of 0.8. The LFM signals are sent to DPMZM1 for CS-SSB modulation to generate a −1st-order optical sideband. A single-tone signal generated from the MSG (Agilent 83630B) is sent to a 90° electrical hybrid coupler (Talent Microwave TBG-20400-3k-90). The outputs of the 90° hybrid are sent to DPMZM2 to realize CS-SSB modulation and generate a +1st-order optical sideband. Following the Dpol-DPMZM, the modulated optical signals in two orthogonal polarization states are combined via PC2 and the polarizer. The combined optical signal from the polarizer is sent to an EDFA (Amonics AEDFA-PA-35-B-FA). The optical signal after amplification from the EDFA is sent to the OBPF (EXFO XTM-50), with a center wavelength of 1549.76 nm and a bandwidth of 56.33 GHz, to reject the out-of-band noise. The optical spectra at the output of the OBPF are shown in Fig. 5(a), where the three separate peaks displayed by blue solid line, red dotted line, and green dashed line are the +1st-order optical sidebands generated by single-tone signals at 10 GHz, 17 GHz, and 25 GHz, respectively. The three overlapping peaks on the right are the −1st-order optical sidebands generated by the LFM signal. In this experiment, the parameters of the LFM signal remain fixed and unchanged. The purple dash-dotted line in Fig. 5(a) displays the frequency response of the OBPF. It is noteworthy that the powers of all optical sidebands shown in Fig. 5(a) are relatively low, which is attributed to an additional 18 dB attenuation of the optical signal output by the OBPF during response measurement. In addition, the optical sidebands of the single-tone signal and the LFM signal appear to have very similar widths in Fig. 5(a), which is mainly due to the resolution limitation of the optical spectrum analyzer (OSA, ANDO AQ6317B) employed for measurements.

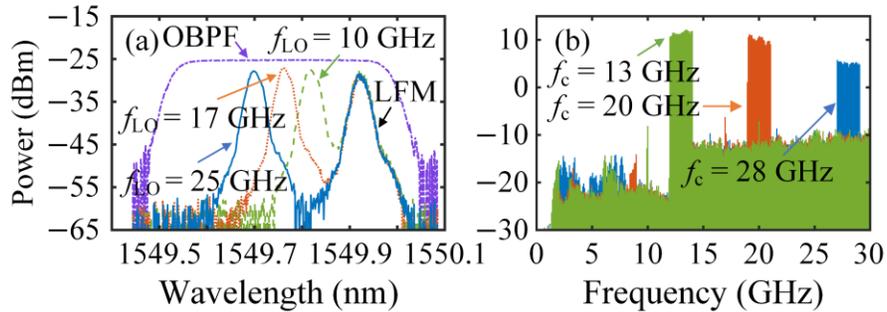

**Fig. 5** (a) Optical spectra at the output of the OBPF and the frequency response of the OBPF. (b) Electrical spectra of the generated LFM signals under different center frequencies.

Then, the optical signal from the OBPF shown in Fig. 5(a) is directed to the PD (u2t MPRV1331A) to generate the frequency-tunable LFM radar signal for simultaneous monitoring human respiration, heartbeat, and speech. The electrical signal generated from the PD is split by the PS (Talent Microwave RS2W20400-K). One output of the PS is amplified by an LNA1 (AT Microwave AT-LNA-0243-5004-LCBT) and then transmitted by the transmitting antenna. The electrical spectra of the transmitted radar signal are measured by an electrical spectrum analyzer (ESA, Rohde & Schwarz FSP-40) and depicted in Fig. 5(b). The green, orange, and blue solid lines are the spectra of the generated 2-GHz bandwidth LFM signals centered at 13 GHz, 20 GHz, and 28 GHz, respectively. It can be seen in Fig. 5(b) that the powers of these LFM signals in these three frequency bands are different, which is mainly due to the lower frequency response of the PD and LNA1 at high frequencies. Thanks to the broadband and flexible characteristics of the microwave photonic system, the frequency of the generated signal can be widely tuned simply by adjusting the frequency of the single-tone signal. While it is theoretically possible to achieve frequency tuning of the generated signal by adjusting the center frequency of the IF LFM signal, this approach is generally avoided due to the complexity associated with generating LFM signals. In addition to adjusting the center frequency, other parameters for the generated LFM signal, such as the sweep bandwidth, need to be controlled by adjusting the relevant parameters of the IF LFM signal. In the following experiments, when we refer to the system operating in the K, Ku, and Ka band, it means that the system uses the aforementioned generated LFM signals with center frequencies of 13 GHz, 20 GHz, and 28 GHz, respectively, as the radar transmission signals.

*3.2 Reconstruction and recognition of speech signals*

In order to evaluate the performance of the proposed radar system in speech detection, we conducted experimental research using the following methods. Firstly, eight volunteers are recruited to record their pronunciation of 20 English words, ranging from "one" to "twenty", using a recording device. Each volunteer repeats the recording of each word's pronunciation 25 times. Subsequently, the aforementioned speech data are played back through high-fidelity loudspeakers. Radar signals generated in Section 3.1 are emitted by the transmitting antenna to detect the diaphragm vibrations of the loudspeaker. Speech can be detected and reconstructed using radar signals in different frequency bands (three in this work). The radar echo signals are received by the receiving antenna and undergo amplification through LNA2 (AT Microwave AT-LNA-0243-4204-LCBT). Following de-chirping processing in the mixer (COM-MW DM8-18-40G-2537), the de-chirped signals are sampled by an OSC (Rohde & Schwarz RTO2032) at a sampling rate of 2.5 MSa/s, which is determined by the frequency of the de-chirped signal in this experiment. Since the sampling rate of the speech signal is equal to the period of the radar signal, under the 100-μs period setting in this work, the effective sampling rate of the speech signal is 10 kSa/s. The acquired waveforms are filtered through a digital domain band-pass filter with a bandwidth of 50 Hz to 2 kHz, followed by spectral subtraction algorithms processing to reconstruct the speech signals. Finally, the reconstructed speech signals are resampled at 8 kHz and segmented into independent speech samples via endpoint detection. In this way, we will obtain 4,000 reconstructed speech samples for each frequency band, totaling 12,000 speech samples across the three bands. In the following sections, a detailed analysis of the reconstructed speech signals is conducted to evaluate the performance of the system.

First, the English word "one" pronounced by a volunteer is reconstructed using our system across three distinct operating frequency bands. The reconstructed results are then compared with the original recording to validate the performance of the proposed system in speech signal recovery. For this purpose, the reconstructed speech signal is resampled to ensure consistency with the data length of the corresponding recorded speech signal. Following this step, cross-correlation-based time alignment is performed to resolve temporal offsets between the signals, after which amplitude scaling is conducted to ensure consistent signal intensity levels. The green, blue, and orange solid lines in Figs. 6(a), 6(c), and 6(e) are waveforms of the reconstructed speech signals "one" of this volunteer in Ku, K, and Ka bands, while the black solid line corresponds to the waveform of the recorded speech signal "one" from the recording device. Comparative analysis illustrates that there are certain differences between the reconstructed and original signals, which may mainly originate from two contributing factors. On one hand, the complex experimental environment may introduce additional interferences. On the other hand, the radar system exhibits poor recovery performance for harmonics above the second order, which also contributes to waveform distortion. As shown in the corresponding spectra presented in Figs. 6(b), 6(d), and 6(f), the power levels of harmonics above the second order in the recorded speech signals exhibit closer proximity to those of the fundamental frequency and second harmonic. In contrast, for the speech signals reconstructed via the radar system proposed in this work, the power of harmonics above the second order is around 20 dB lower than that of the fundamental and second-order harmonics. Nevertheless, based on the spectral evolution trends, it can be observed that across the three distinct frequency bands employed in the experiments, the reconstructed speech signals exhibit remarkably high similarity to the originally recorded speech spectra.

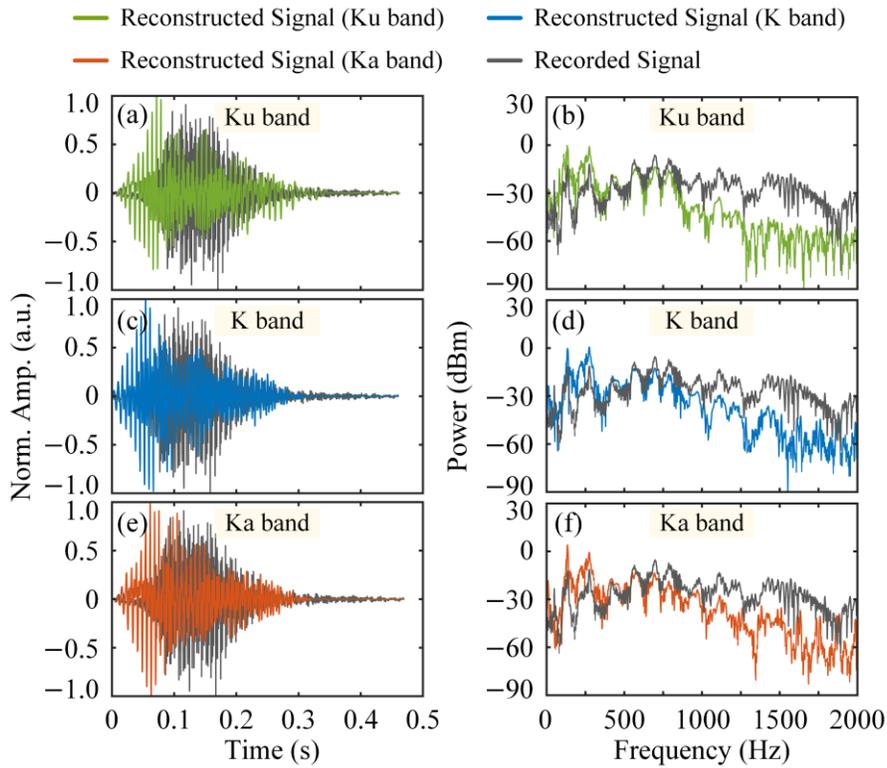

**Fig. 6** Comparison of reconstructed and recorded speech signals of "one": Ku band (a) waveforms and (b) spectra; K band (c) waveforms and (d) spectra; Ka band (e) waveforms and (f) spectra.

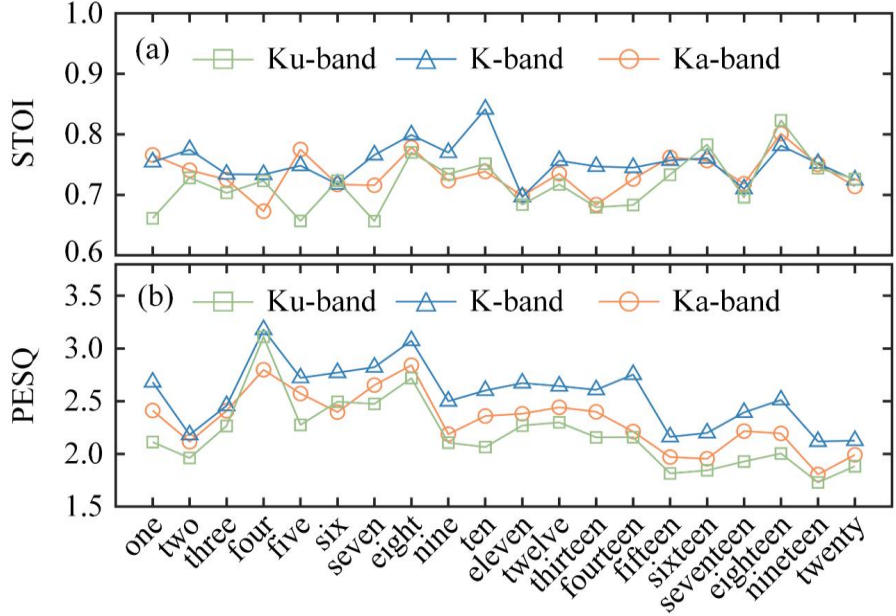

**Fig. 7** (a) STOI scores and (b) PESQ scores of the reconstructed speech signals in Ku, K, and Ka bands.

To quantitatively and comprehensively evaluate the quality of reconstructed speech signals, we employ two objective metrics: Perceptual Evaluation of Speech Quality (PESQ)[33] and Short-Time Objective Intelligibility (STOI)[34], to analyze the 12,000 reconstructed speech samples from the eight volunteers. The STOI metric yields scores within the range [0, 1], while PESQ produces values spanning [−0.5, 4.5]. In both cases, higher scores indicate superior speech reconstruction performance. Figure 7(a) shows the average STOI scores of reconstructed speech signals from "one" to "twenty" on three different frequency bands, all of which are above 0.65. To better evaluate the performance of the reconstructed speech signals in the three frequency bands, the average STOI scores for all words in each band are further calculated. It can be seen that the STOI scores for Ku, K, and Ka bands are 0.7192, 0.7536, and 0.7351, respectively. These results indicate reconstructed speech signals in all three bands can be well understood. Additionally, as illustrated in Fig. 7(b), the PESQ scores for each word across the three distinct frequency bands have also been calculated, revealing average values exceeding 1.7. Similarly, when averaging the PESQ scores for all words within each frequency band, the results showcase values of 2.1841 for the Ku band, 2.5592 for the K band, and 2.3158 for the Ka band. These findings further corroborate that the proposed method is capable of reconstructing speech signals with exceptional quality.

The reconstructed speech signals are further processed through three distinct models: two single channel baseline models utilizing the spectrogram and the Mel-spectrogram as single feature input, and a dual-channel feature fusion CNN model. Within each of the three models, the reconstructed speech signals in the Ku, K, and Ka bands were structured into datasets comprising 4000 segments per band. These segments were proportionally divided across all bands: 2,800 for training, 600 for validation, and the remaining 600 for testing. To ensure objective comparison of recognition accuracy across the models, the speech signals in all three bands underwent five training iterations per model, with average accuracy computed from the results. As depicted in the histogram presented in Fig. 8, the recognition accuracies for the majority of words across all three frequency bands surpass 90% in the three models. Notably, for most words, the recognition accuracies achieved by the dual-channel feature fusion model are higher than that of the single-channel baseline models. The baseline model that employs the spectrogram yields average recognition accuracies of 95.30%, 95.73%, and 95.53% for the Ku, K, and Ka bands, respectively. Meanwhile, the model utilizing the Mel-spectrogram achieves higher average recognition accuracies, reaching 96.50%, 96.87%, and 96.70% for the corresponding bands. In stark contrast, after undergoing training with the two-channel feature fusion model, there is a marked enhancement in the average recognition accuracies across all three bands. Notably, when employing the two-channel feature fusion model, the average recognition accuracy for reconstructed speech signals in the K band soars to 98.07%, surpassing the spectrogram-based model by 2.34% and the Mel-spectrogram-based model by 1.2%. Furthermore, the average recognition accuracies for the reconstructed speech signals in the Ku and Ka bands register at 97.20% and 97.43%, respectively, also exhibiting improvements over the two single-feature baseline models. These results confirm the effectiveness of multimodal feature fusion in enhancing speech recognition accuracy. There are certain differences in speech recognition accuracy across the three frequency bands, which are primarily attributed to the varying degrees of coupling between the same acoustic vibration characteristics and different radar frequencies. However, these differences are not significant. We can also conclude that, by leveraging the broadband tunability of microwave photonic radar, accurate recovery and recognition of speech signals can be achieved across different operating frequency bands.

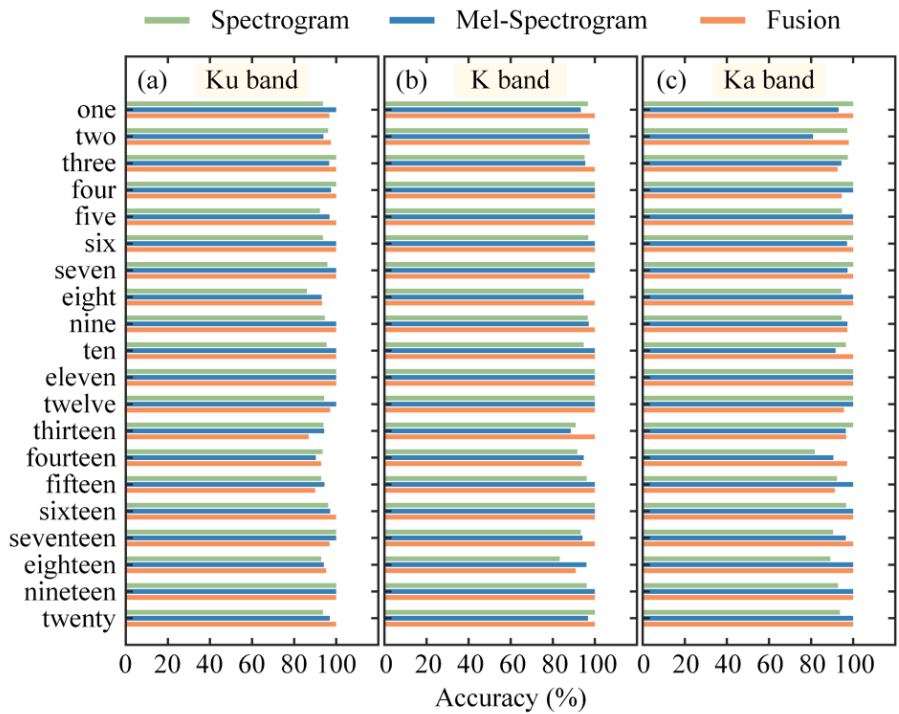

**Fig. 8** Recognition accuracy for reconstructed speech signals in (a) Ku, (b) K, and (c) Ka bands by using Spectrogram-based model, Mel-spectrogram-based model, and the dual-channel feature fusion model.

## 3.3 Concurrent vital sign monitoring and voice detection

Notably, the microwave photonic radar system introduced in this work is not only capable of detecting speech signals but also able to simultaneously monitor weak vibration information induced by physiological activities such as human respiration and heartbeat. In this experiment, a respiratory and heartbeat simulator composed of two balloons and their control devices is employed. As shown in Fig. 9, to replicate the subtle vibrations associated with human respiration and heartbeat, two balloons are securely fastened to metal posts and arranged at an identical radial distance from the antenna as the loudspeaker—specifically, positioned roughly 1 meter in front of the antenna pair. Balloon 1 is connected to an electric pump, which inflates and deflates the balloon at a lower frequency to simulate human respiration. Balloon 2 is inflated and deflated by a manual pump at a relatively higher frequency to simulate human heartbeat. The radar system simultaneously detects both speech signals and vital signs across the aforementioned three distinct frequency bands.

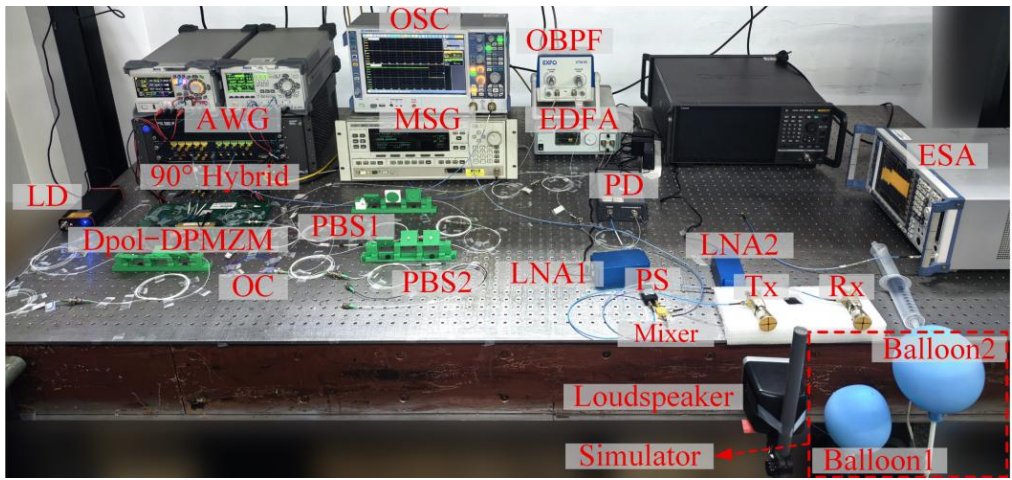

**Fig. 9** Experimental setup of simultaneous vital sign monitoring and speech detection.

The echo signals reflected by the balloons and the diaphragm of the loudspeaker are received by the receiving antenna and sampled by the OSC after de-chirping. Due to the period of the radar pulse signal remaining unchanged, the sampling rate of the weak vibration signal is still 10 kSa/s. The data acquisition duration is 20 seconds. After digital signal processing, the speech, respiration, and heartbeat signals at different frequencies are separated by different bandpass filters. To achieve the extraction of respiration and heartbeat data, two bandpass filters with passbands from 0.13 to 0.5 Hz and from 0.8 to 1.9 Hz are used, respectively, and the respiration and heartbeat monitoring results are shown in Fig. 10. Figures 10(a) and 10(d) depict the respiration waveform of the Ku band de-chirped radar signal after the bandpass filtering and its corresponding spectrum after FFT, while Figs. 10(g) and 10(j) show the heartbeat waveform after the bandpass filtering and its corresponding spectrum of the Ku-band de-chirped radar signal. The spectral analysis reveals that the frequency peaks corresponding to respiration and heartbeat exhibit significantly higher amplitudes than the noise floor, thereby validating the system's feasibility for vital sign detection. The results of K and Ka bands are also shown in Fig. 10, which is largely similar to those of Ku band. The peak frequencies in the spectra represent the respiratory and heartbeat rates. It should be noted that the frequency values corresponding to the results vary across different frequency bands, as these several measurements were conducted at different times.

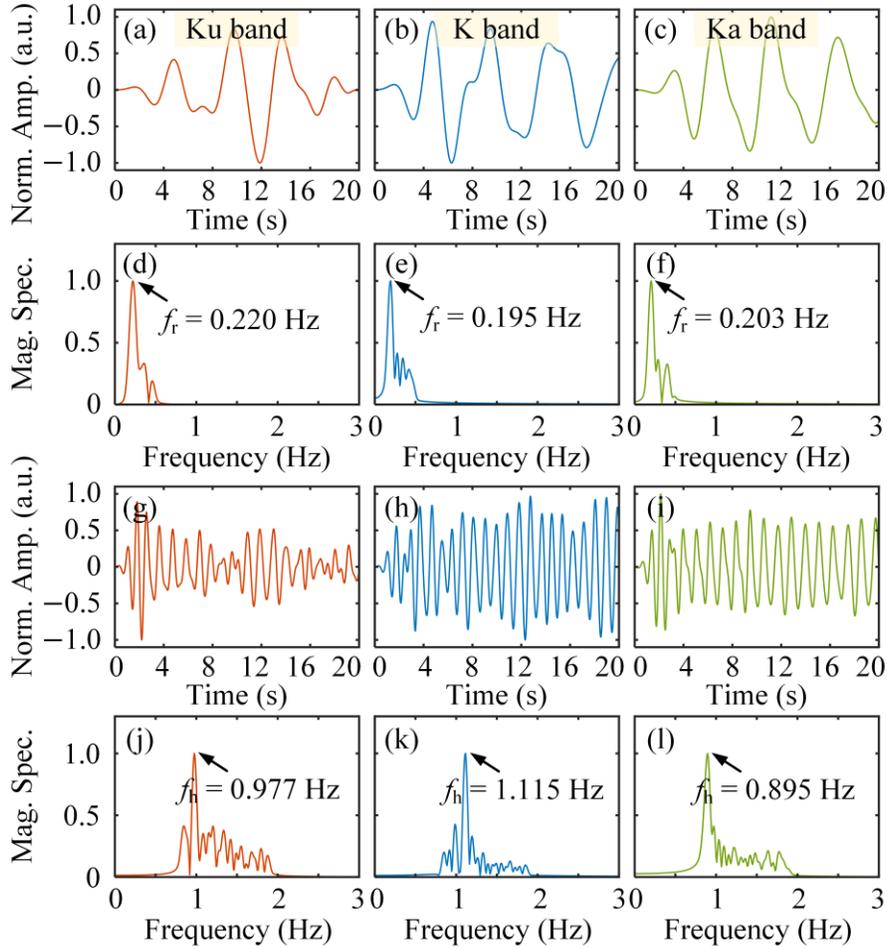

**Fig. 10** Waveforms of respiration in (a) Ku, (b) K, and (c) Ka bands. Waveforms of heartbeat in (g) Ku, (h) K, and (i) Ka bands. Spectra of respiration in (d) Ku, (e) K, and (f) Ka bands. Spectra of heartbeat in (j) Ku, (j) K, and (l) Ka bands.

The peak frequencies of the respiration and heartbeat spectra in Fig. 10 are converted into the corresponding counts in 20 seconds, which are given in Table 1. Specifically, the monitored counts are derived from frequency-domain signal processing, whereas the actual counts represent manually verified ground-truth measurements. The maximum absolute error counts of the respiration and heartbeat in 20 seconds are 0.40 and 2.46 in the three frequency bands, respectively. The error count in heartbeat detection exceeds that of respiration, primarily due to differences in the stability of the balloon actuation mechanisms. Specifically, the respiration-simulating balloon is driven by an electric pump with a precisely controlled frequency, ensuring consistent inflation-deflation cycles. In contrast, the heartbeat-simulating balloon is manually operated, resulting in inherent variability in the cycle rate that contributes to higher detection errors.

Table 1. Comparison of monitored and actual counts of respiration and heartbeat in 20 seconds.

|  |  | Ku band | K band | Ka band |
|---|---|---|---|---|
| Respiration | Monitored count | 4.40 | 3.90 | 4.06 |
|  | Actual count | 4.00 | 4.00 | 4.00 |
|  | Error count | 0.40 | −0.10 | 0.06 |
| Heartbeat | Monitored count | 19.54 | 22.30 | 17.90 |
|  | Actual count | 22.00 | 24.00 | 18.50 |
|  | Error count | −2.46 | −1.70 | −0.60 |

To better demonstrate the accuracy of the proposed system in monitoring respiratory and heart rates, ten measurements are further conducted across the three frequency bands. The monitored counts, actual counts, and error counts of respiration and heartbeat in 20 seconds are shown in Fig. 11. Figure 11(a) shows the measurement results of the Ku band. In this figure, the solid and dashed lines in blue (representing respiration) and orange (representing heartbeat) denote the ground-truth and system-monitored counts, respectively. The green and purple dotted lines denote the corresponding absolute error counts for respiration and heartbeat. Figures 11(b) and 11(c) present the measurement results obtained in the K and Ka bands, respectively, with line color and style conventions consistent with those in Fig. 11(a). The average absolute error counts of the respiration in 20 seconds are obtained as 0.31 (Ku band), 0.35 (K band), and 0.39 (Ka band). The maximum absolute error counts of the respiration are 0.72 (Ku band), 0.88 (K band), and 0.88 (Ka band). The average absolute error counts of the heartbeat in 20 seconds are 0.62 (Ku band), 0.87 (K band), and 0.42 (Ka band). The maximum absolute error counts of the heartbeat are 2.46 (Ku band), 2.02 (K band), and 0.74 (Ka band).

Quantitative error analysis reveals that the respiration and heartbeat counts can be accurately monitored in three different frequency bands.

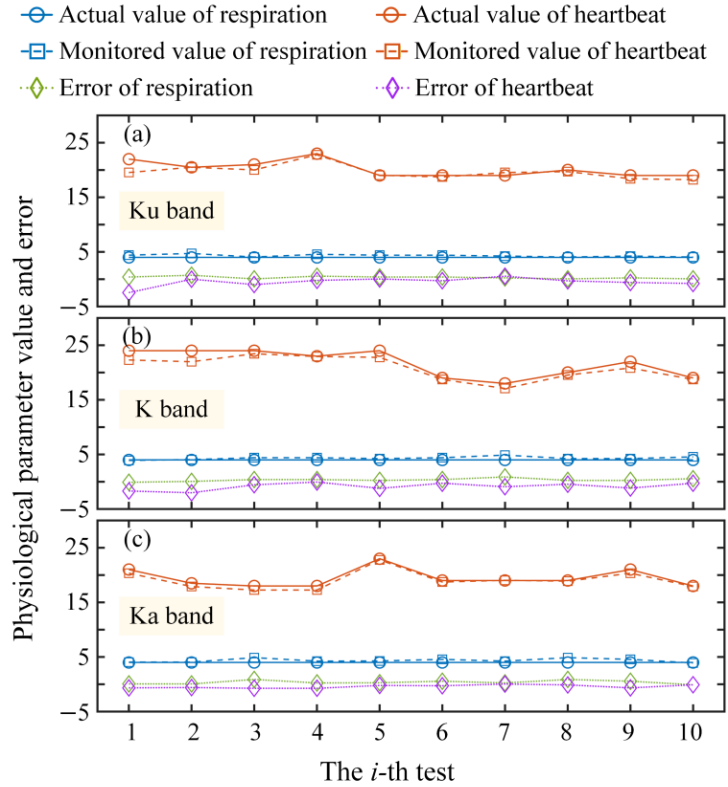

**Fig. 11** Monitored counts, actual counts, and error counts of respiration and heartbeat in (a) Ku, (b) K, and (c) Ka bands.

To demonstrate the capability of the system for simultaneous detection of respiration, heartbeat, and speech, the acquired waveforms undergo further processing. The bandwidth of the digital bandpass filter used to extract the speech signal is set from 50 Hz to 2 kHz. The waveform of the reconstructed speech signals from English word "one" pronounced multiple times in the Ku band are depicted in Fig. 12(a), with the insets showing enlarged views of the first and eighth segments. Figures 12(a-i) and 12(a-iii) present the spectrogram and Mel-spectrogram for the first segment of the reconstructed speech signal, while Figs. 12(a-ii) and 12(a-iv) show the spectrogram and Mel-spectrogram for the eighth segment. Compared to the spectrogram, the Mel-spectrogram exhibits superior frequency resolution at low frequencies, enabling more effective capture of detailed variations in these regions. The results for the K and Ka bands are also shown in Figs. 12(b) and 12(c). When using a speaker to play back the reconstructed speech signals in these three bands, the speech contents can be clearly heard. Furthermore, the reconstructed speech signals across three distinct frequency bands are also subjected to analysis, yielding STOI and PESQ scores. The average STOI scores for the reconstructed speech signals in the Ku, K, and Ka bands are 0.7858, 0.6997, and 0.8165, respectively, while the corresponding average PESQ values are 2.3440, 2.1886, and 2.5653, respectively. The results indicate that the system can not only accurately monitor the respiration and heartbeat rate but also reconstruct the voice signals well.

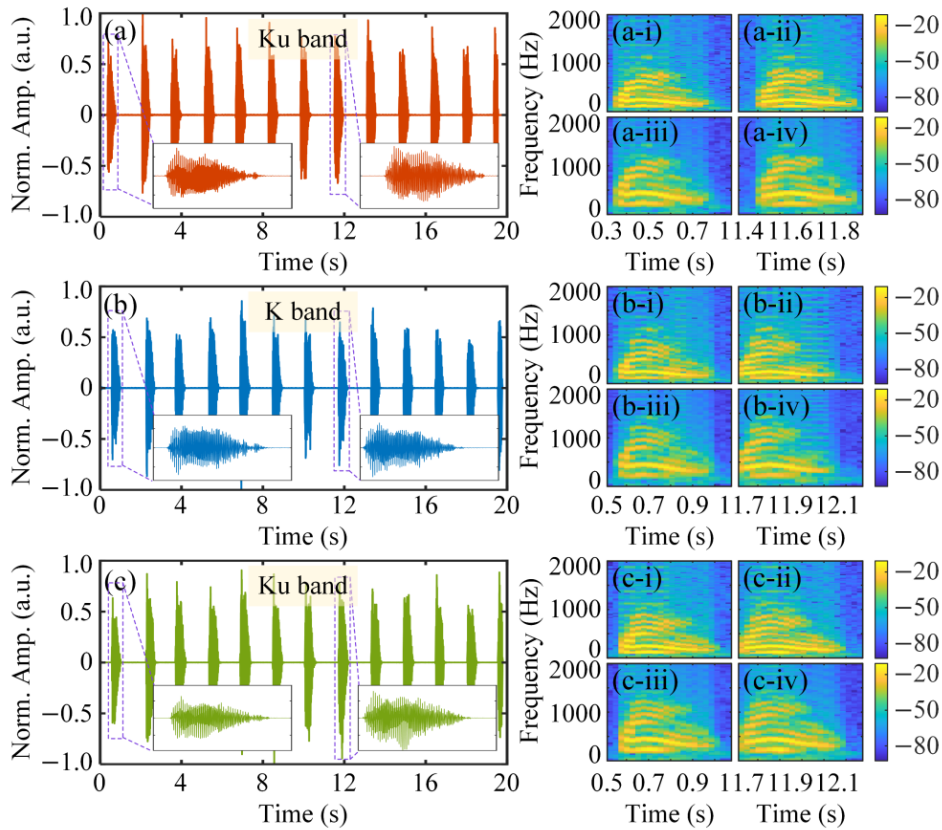

**Fig. 12** Waveforms of the reconstructed speech signals at 20 seconds in (a) Ku, (b) K, and (c) Ka bands. (i) and (ii) are the corresponding spectrograms of the first and eighth speech segments. (iii) and (iv) are the corresponding Mel-spectrograms of the first and eighth speech segments.

*3.4 Discussion*

In the above study, we have validated the simultaneous detection of human speech signals as well as vital signs such as respiration and heartbeat across the Ku, K, and Ka bands using a microwave photonic radar system. Notably, we employ a dual-channel feature fusion CNN model to achieve high-precision recognition of speech signals. Moving forward, we will now explore the advantages and significant implications of applying broadband tunable microwave photonic radar to the detection of human speech signals and vital signs. In short, the broadband tunable microwave photonic radar demonstrates superior adaptability and robustness when applied to the aforementioned detection tasks, especially the speech detection task.

For example, when detecting weakly vibrating targets, such as when a person is speaking softly or producing faint sounds, utilizing radar signals in higher frequency bands for detection can convert subtle vibrations into more pronounced phase changes, thereby enabling the system to reconstruct minute vibrations with greater accuracy. However, during long-distance target detection or when penetrating obstacles, high-frequency radar signals suffer from higher losses and attenuation, which is counterproductive for monitoring the vibration characteristics of the target. Therefore, when the system is designed for detecting distant targets, switching to lower radar frequency bands yields better detection results. Since the amplitude of chest vibrations caused by respiration and heartbeat is typically much greater than the vibrations of the laryngeal skin induced by human vocalization, in the following experiment, we take speech detection as an example to further demonstrate the performance of the microwave photonic radar system under different scenarios.

First, we investigate the detection capabilities of radar signals at different frequency bands for varying vocal intensities. To verify the impact of vocal intensity on the quality of the reconstructed signal, we keep the position of the loudspeaker relative to the antenna pair fixed. Subsequently, we have a volunteer pronounce the word "one" twelve times at each of the three volume levels: 100%, 50%, and 25%. To streamline the experiment and observe more intuitively the quality of the reconstructed speech signals at different frequencies under varying vocal intensities, we will only consider the low-frequency (Ku band) and high-frequency (Ka band) bands among the aforementioned three frequency bands. The LFM radar signals centered at 13 GHz and 28 GHz, with a 2-GHz bandwidth are emitted from the transmitting antenna, and the reflected echo signals are de-chirped, followed by data acquisition via the OSC. Two segments of the reconstructed speech signals after signal processing are shown in Fig. 13.

Figure 13(a) presents the unwrapped phase waveforms of the reconstructed speech signals in the Ku band under three different volume levels. The reconstructed speech waveforms under different volume levels are offset in time to facilitate better presentation on the same figure. It is evident that as the volume decreases, the amplitude of the reconstructed speech signal drops significantly. In particular, when the volume is set to 25%, the amplitude of the reconstructed signal is much smaller than that when the volume is at 100% and 50%. Figure 13(b) shows the corresponding results in the Ka band. Notably, the reconstructed speech signals in the Ka band exhibit significant larger amplitudes than those in the Ku band at the same volume level, which is attributed to the higher frequency of the Ka band making it more sensitive to phase changes. Additionally, it should be worthy mentioned that although, in the Ka band, the amplitude of the reconstructed speech signal at 25% volume (purple rectangles in Fig. 13(b)) is also much smaller than that at 100% and 50% volumes, compared with the reconstructed results at the same volume levels in the Ku band, it demonstrates a distinct higher amplitude, as illustrated by the corresponding comparison in Fig. 13(c). By further using a speaker to play back the waveforms of the reconstructed speech signals, when the volume is reduced to 25%, the waveform reconstructed in the Ku band has such a small amplitude that no effective speech information can be audibly perceived. In contrast, relatively clear speech information can still be heard by playing back the waveform of the speech signal reconstructed in the Ka band.

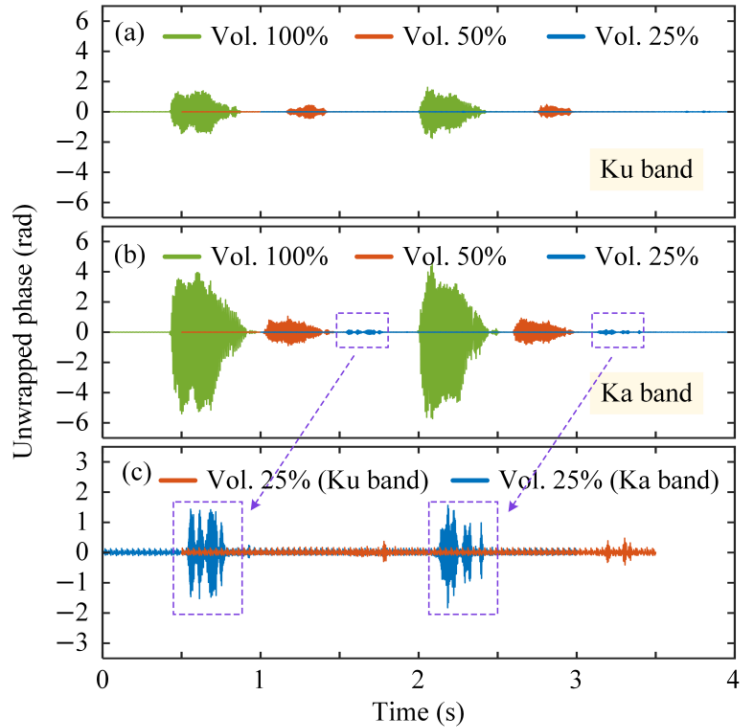

**Fig. 13** Waveforms of reconstructed speech signals at different volumes in (a) Ku band and (b) Ka band. (c) Enlarged waveforms of reconstructed speech signals at 25% volume in the Ku and Ka bands.

To conduct a thorough and quantitative analysis of how volume changes impact the quality of speech signals, Table 2 presents the average scores of STOI and PESQ for all twelve segments of speech signals across each frequency band at the above three volume levels. The metrics clearly demonstrate that in both the Ku and Ka bands, a reduction in volume results in a substantial decline in the quality and intelligibility of the reconstructed speech signals. In terms of quantitative metrics, for the Ku-band, STOI and PESQ scores cannot be calculated when the volume is set to 25%, as it becomes virtually impossible to retrieve any meaningful speech information under such conditions. However, when the volume is adjusted to 25%, the STOI and PESQ scores of the reconstructed speech signals in the Ka band reaches 0.5790 and 1.2952, respectively. Although these scores indicate relatively poor performance, under such STOI and PESQ levels, the speech signals can still retain a certain degree of intelligibility. The results fully demonstrate that the Ka band outperforms the Ku band in recovering speech signals under low-volume scenarios, which is crucial for recovering weak speech signals. The broadband tunability of the proposed microwave photonic radar allows for flexible switching to high-frequency bands in corresponding scenarios, thereby achieving superior speech signal reconstruction performance.

**Table 2.** Average scores of STOI and PESQ of speech signals at different volumes in Ku and Ka bands.

| Volume | Ku band | | Ka band | |
|---|---|---|---|---|
| | STOI | PESQ | STOI | PESQ |
| 100% | 0.7547 | 2.0080 | 0.8228 | 2.7606 |
| 50% | 0.6860 | 1.5844 | 0.6968 | 1.8423 |
| 25% | / | / | 0.5790 | 1.2952 |

Then, we aim to further verify the impact of radar signals in different frequency bands on the quality of speech signal recovery under varying target distances. Due to the limited space of our laboratory, we are unable to adjust the target distance over a wide range. Therefore, in the experiment, we adopted the following equivalent approach: we changed the output power of the radar signal by adjusting the optical power input to the PD. Consequently, the radar echo power at the receiving end would change in response, which could approximately simulate the effect of target distance on the radar echo. In the experiment, a recording of a volunteer repeatedly reading the word "one" was played back through the loudspeaker. The volume was set to 100%, and the recording was detected using radar signals in the Ku and Ka bands, respectively.

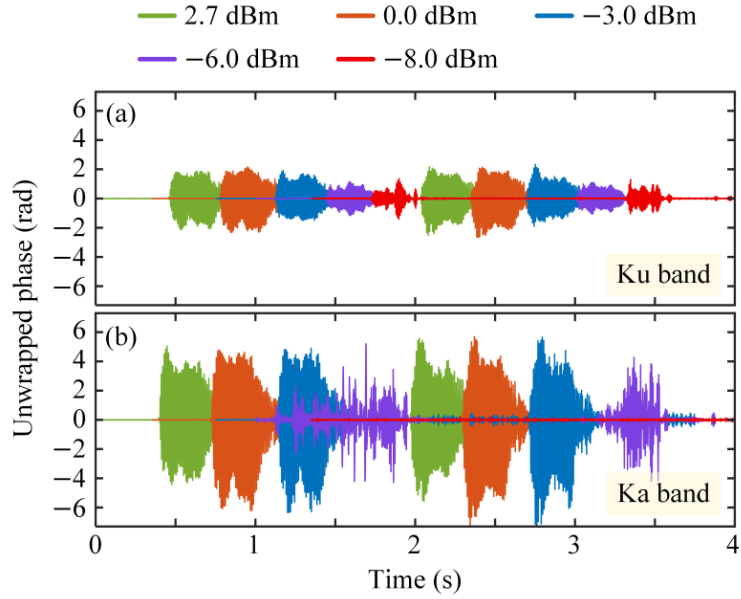

**Fig. 14** Waveforms of reconstructed speech signals at different powers in the (a) Ku and (b) Ka bands.

**Table 3.** STOI and PESQ scores of speech signals at different volumes in Ku and Ka bands.

| Power | Ku band | | Ka band | |
| --- | --- | --- | --- | --- |
| (dBm) | STOI | PESQ | STOI | PESQ |
| 2.7 | 0.7554 | 2.4868 | 0.8211 | 2.6703 |
| −3.0 | 0.6874 | 2.1207 | 0.7350 | 2.4044 |
| −8.0 | 0.6411 | 1.7166 | / | / |

Figure 14 presents the waveforms of the reconstructed speech signals in the Ku and Ka bands when the optical power input to the PD is set to 2.7 dBm, 0.0 dBm, −3.0 dBm, −6.0 dBm, and −8.0 dBm, respectively. The reconstructed speech waveforms are also offset in time to facilitate better presentation on the figure. The results demonstrate that high-quality speech signal reconstruction can be successfully accomplished in both the Ku and Ka bands, as long as the optical power fed into the PD remains above −3.0 dBm. It can also be observed that when the power is above −3.0 dBm, the amplitude of the signal reconstructed in the Ka band is greater than that in the Ku band. This is similar to the results shown in Fig. 13, primarily because high-frequency signals possess better capabilities for detecting weak vibrations. Once the power drops to the level of −6.0 dBm, a noticeable performance gap emerges between these two bands. Although it can be observed that the waveform reconstructed in the Ka band exhibits a relatively large amplitude, when compared with the waveform reconstructed in the Ku band, its regularity is noticeably poorer and it contains a significant amount of noise. When playing back the reconstructed waveforms, the speech signal reconstructed from the Ku band can be clearly heard and understood. In contrast, although the speech signal reconstructed from the Ka band can also be understood, it contains very pronounced noise. Once the optical power input to the PD is further reduced to the critical level of −8.0 dBm, the performance gap between the two bands becomes even more pronounced. Utilizing radar signals in the Ka band for speech reconstruction exhibits a very small amplitude, as it barely succeeds in restoring any meaningful amount of effective speech information. By contrast, the speech signal can still be effectively restored in the Ku band. Although the amplitude of the reconstructed signal is decreased, the speech content remains clearly audible. Table 3 lists the average STOI and PESQ scores of the reconstructed speech signals in the above experiment. It can be clearly observed that when the microwave photonic radar operates in the Ku band and the optical power input to the PD is −8.0 dBm, the STOI and PESQ scores of the reconstructed speech signal are 0.6411 and 1.7166, respectively. Although these values indicate a generally moderate signal quality, they still signify that the speech signal retains a certain level of speech quality. In contrast, due to the lack of effective speech information, corresponding scores cannot be calculated in the Ka band. This result fully confirms that in scenarios with low signal power (equivalent to a long target distance), the speech reconstruction performance in the Ku band is significantly superior to that in the Ka band. This is primarily because the free-space loss of Ku band signals is smaller, and the same broadband antenna exhibits greater antenna gain at the Ku band.

By synthesizing the results presented in Figs. 13 and 14 as well as Tables 2 and 3, we can conclude the following for practical applications: When conducting weak speech and vibration detection, we can adjust the operating frequency band of the microwave photonic radar to a higher frequency range to achieve better speech and vibration reconstruction results. On the other hand, when detecting weak speech and vibrations from long-distance targets, lower frequency bands often provide superior reconstruction performance, depending on the actual distance of the target. In practical applications, it is necessary to further explore and select an appropriate operating frequency band based on the target's proximity and the intensity of vibrations to achieve the optimal reconstruction results.

## 4  Conclusion

In summary, we have proposed a broadband tunable microwave photonic radar system capable of simultaneously monitoring multimodal vital signs and detecting human speech signals. In addition to accurately measuring human respiratory and heartbeat rate using the microwave photonic radar system, in the realm of speech detection, high-quality restoration and recognition of human speech signals has been achieved through the collaborative operation of speech signal reconstruction based on microwave photonic radar and a convolutional neural network that incorporates dual-channel feature fusion using spectrogram and Mel-spectrogram. This approach has significantly enhanced the accuracy and robustness of speech recognition. Experimental results demonstrate that, under default settings (100% volume, 1 m target distance, and 2.7-dBm PD input power), the speech recognition accuracies for the Ku, K, and Ka bands are 97.20%, 98.07%, and 97.43%, respectively. Furthermore, the maximum average error counts for respiratory and heartbeat monitoring over a 20-second period across the three frequency bands are 0.39 and 0.87, respectively, indicating a high level of monitoring accuracy. Additionally, benefiting from the wideband tunability of the microwave photonic system, we conducted speech recognition tests under varying volume levels and equivalent distances. The results show that the system can flexibly select the optimal operating frequency band according to actual application scenarios, thereby achieving optimal performance. This capability is expected to greatly enhance the system's adaptability to a wide range of application scenarios. To the best of our knowledge, this study is the first to integrate broadband microwave photonic radar with respiratory and heartbeat monitoring as well as speech signal detection and recognition. The proposed system realizes an innovative application of multimodal physiological information detection and exhibits significant advantages in high-accuracy monitoring and frequency band adaptability. In the future, with the continuous development and maturation of cutting-edge technologies such as integrated photonic chips and their deep integration with this system, the proposed solution has the potential to become a key technology in the field of multimodal physiological monitoring, providing crucial technical support for areas such as medical health, public safety, etc.

*Disclosures*

The authors declare no conflicts of interest.

*Acknowledgments*

National Natural Science Foundation of China (62371191, 62401207); Shanghai Oriental Talent Program (QNJY2024007), Science and Technology Commission of Shanghai Municipality (22DZ2229004).

*Code, Data, and Materials Availability*

Data underlying the results presented in this paper are not publicly available at this time but may be obtained from the authors upon reasonable request.